# The Epistemic Risk of the 31st Spore: If Planets Aren't Fine Tuned, We're Doing Life Detection Wrong

Harrison B. Smith[1,2,*,✣] and Cole Mathis[3,4,⁎,✣]

[1]Earth-Life Science Institute, Institute of Science Tokyo, Ookayama, Meguro-ku, Tokyo, Japan
[2]Blue Marble Space, Seattle, Washington, USA
[3]School of Complex Adaptive Systems, Arizona State University, Tempe, AZ, USA
[4]Biodesign Institute, Arizona State University, Tempe, AZ, USA
* hbs@elsi.jp
⁎ cole.mathis@asu.edu
✣Equal Contribution

## Abstract

Exploration of planetary bodies within our solar system will be essential for understanding the origin of life on Earth and the distribution of life in the universe. Planetary protection policy is concerned with balancing this desire for exploration against the risks of contaminating alien planets with Earth life, and contaminating Earth with alien life. However, at present, we have no fundamental scientific understanding of life's emergence or its nature beyond Earth. Given this nearly complete ignorance about the possibility of alien life, or Earth life's capacity to expand beyond our planet, it is difficult to reason about the real risks of space exploration. Here we contend that contemporary understandings of the risks of forward contamination are based on arguments which inconsistently apply our incomplete knowledge to the problem. We reason that a more assertive posture towards space exploration, focused on determining whether other planets in the solar system are inhabited, is warranted and explain why such a posture may not increase the epistemic risks of planetary contamination. Finally, we explore the consequences of our arguments for planetary protection protocols, and life detection efforts.

## Prologue

Planetary catastrophe sits on a razor's edge. The cause? A single errant microbe, sitting in the wrong place, at the wrong time. A self-contained weapon of mass-destruction with the capacity to germinate global extinction events, forever preventing humanity from acquiring fundamental knowledge about the universe. This microbe is not just any microbe, it is the 31st of its kind. It originates, as the most ardent conspiracy theorist might suppose, in a highly secure government facility. There government agents, using the full force of modern science, attempted to kill its ancestors and failed, and in the process made this microbe more resilient to eradication. But—we humans need not worry. This microbe will not be the cause of our next global pandemic. Because this microbe is not destined for our globe.

Microbe 31 rides innocuously alongside its companions on an interplanetary spaceship launched from Earth. It does not look remarkably different from the other 30 microbial stowaways, because it is not remarkably different from them. Yet, its very presence is a catastrophe. It sits, like the other microbes, on the surface of a rover bound for Mars. If this were most any other spacecraft, microbe 31 would not be worth talking about. In fact, if this were most any other spacecraft, microbe 31 would be just another brick in the wall—one of thousands, or tens of thousands, of spores that are routinely allowed to hitch a ride to the red planet. But, this is not just another spacecraft to Mars—not just another mission to rediscover ancient evidence of liquid water which has long since escaped into space or become bound in minerals. The mission of *this* spacecraft is to look for living, *metabolizing*, Martian life.

However, the existence of microbe 31 renders this an impossible task. Not only does microbe 31 interfere with our ability to detect and characterize potential native Martian life, but it poses a risk for that life's very existence, and perhaps for the existence of the Martian biosphere as a whole.

Why does microbe 31 pose such perilous risk to Mars? And is that risk justified in preventing our use of the most decisive and obvious strategies to search for extraterrestrial life?

## The Purported Risks of Forward Contamination

In the 1950s and early 1960s, on the heels of the launch of Sputnik, serious discussions began on the idea of planetary contamination [1–6]. Concerns were identified in both the direction of forwards (Earth life contaminating an extraterrestrial body), and backwards (extraterrestrial life contaminating Earth) contamination. These concerns began ongoing deliberation first at the national scale (via the US National Academy of Sciences), and then through various bodies at the international scale, eventually ending up under the purview of the Committee on Space Research (COSPAR, where it remains today) [7]. In 1964 COSPAR issued Resolution 26 (COSPAR, 1964, p. 26), which affirms that [8]:

> The search for extraterrestrial life is an important objective of space research, that the planet of Mars may offer the only feasible opportunity to conduct this search during the foreseeable future, that contamination of this planet would make such a search far more difficult and possibly even prevent for all time an unequivocal result, that all practical steps should be taken to ensure that Mars be not biologically contaminated until such time as this search can have been satisfactorily carried out, and that cooperation in proper scheduling of experiments and use of adequate spacecraft sterilization techniques is required on the part of all deep space probe launching authorities to avoid such contamination.

The Nobel Prize winner Joshua Lederberg, another early influential voice on the topic had also written [3]:

> The introduction of microbial life to a previously barren planet, or to one occupied by a less well-adapted form of life, could result in the explosive growth of the implant, with consequences of geochemical scope. With a generation time of 30 minutes and easy

> dissemination by winds and currents, common bacteria could occupy a nutrient medium the size of the earth in a few days or weeks, being limited only by the exhaustion of available nutrients.

These are remarkably consequential postulations. Since the 1960s, we have of course learned much about biology, planetology, and the constitution of our own solar system. While Mars may still be the only Earth-like planet in the solar system that could, in-principle, currently support life as we know it[1], few scientists today would argue that it offers "the only feasible opportunity" to search for life in our solar system [9]. Yet, the beliefs and philosophy central to these historical documents have remained remarkably calcified—cheif among them that in-situ searches for life could jeopardize the existence of the very life we hope to detect.

Consider the increasingly stringent bioburden requirements the closer a spacecraft comes to so-called "special regions" thought to be hospitable to extant life [10]. For example, spacecraft landing in a "not special" region of Mars, carrying instruments unrelated to investigations of extant Mars life, is restricted to a surface bioburden level of $\leq 3 \times 10^5$ spores[2]; but a spacecraft landing anywhere on Mars, carrying instruments related to investigations of extant Mars life, is restricted to a surface bioburden level of ≤ 30 spores. Why 30 spores? Because current bioburden requirements are defined relative to the sterilization level used on the Viking spacecraft, which "begin with $N = 3 \times 10^5$ spores allowed (Viking pre-sterilization level) and apply a sterilization process to reduce the total N by 4 orders of magnitude—equivalent to 30 surface spores."[8]. However these thresholds are not theoretically or empirically motivated[3]. In fact, they were established prior to a modern understanding of microbial evolution that practically considers ecological resistance, propagule pressure, stochasticity, and abiotic factors in regulating invasive potential of novel communities [11–15].

We start by identifying that there exist both moral and epistemic risks of planetary contamination, focusing our attention on the epistemic risks. This leads to an examination of the logic underpinning the fear reflexively evoked by planetary contamination, and its potential effect on our ability to detect native[4] extraterrestrial life. Our examination of this logic focuses on three angles: the likelihood of abiogenesis, the perceived fine-tuning of biosphere stability, and the potential for biological invasion (and thus contamination) between distinct biospheres. Our analysis leads us to conclude that, even with (and perhaps because of) our complete ignorance on the likelihood of abiogenesis, it is paradoxical to worry that forward contamination would impact our ability to detect life outside Earth. The same is true for the risks of accidental destruction of extra-terrestrial biospheres, especially in the context of existing terrestrial contamination on bodies like Mars. We end by exploring the practical consequences of our conclusion—that it changes how we should look for non-Earth life both on other planets and our own, and how it radically changes how we think about spacecraft sterilization.

---

[1] While Venus is also an Earth-sized rocky planet with an atmosphere, we exclude it here because its conditions such as high sulfuric acid concentration and low water activity are incompatible with any known Earth life [66].

[2] We use spore and microbe interchangeably, although they have different technical definitions. In general, our language is meant to be consistent with documents such as COSPAR [10], which say e.g., that spores are "aerobic microorganisms that survive a heat shock of 80°C for 15 minutes".

[3] At the time of their estimation these parameters were a source of intense debate, see for example [4,71].

[4] What constitutes "native" life can get a bit murky because a biosphere on Mars could be distinct from Earth's biosphere (having undergone its own abiogenesis), or it could share an origin with Terran life (having established itself on Mars after natural material exchange). While much ink has been spilled hammering out these concepts as they relate to ecological dynamics on Earth (see e.g., [14,15,59], a deeper discussion on how these apply to interplanetary invasion warrants its own paper. Here, by native life, we mean life which is not descended from any organisms transported by human spacecraft.

## Epistemic Risks Motivate Planetary Protection

It's worth discussing *why* we might care about protecting extra-terrestrial planets from Earth life. We can separate the risks into two categories: moral and epistemic. Moral risks relate to whether our actions could lead to outcomes that are not aligned with human/social values, which may or may not be universally held. For example, is it wrong to "contaminate" a planet, whether living or not? Is it wrong if native extra-terrestrial life is harmed in our quest to detect it? On the other hand, epistemic risks relate to how our actions could facilitate or inhibit our ability to gain knowledge about the world. For example, is there a risk that contamination could prevent detection? These two categories of risk are not independent. For instance, perhaps we have the moral obligation to preserve the potential collection of knowledge for future generations; or maybe it's morally wrong to disturb the non-human causal order of systems, regardless of whether those systems comprise living processes. However, our moral obligations to non-human species, non-Earth abiotic environments, and non-Earth life are all potentially fraught conceptual issues [16,17]. Here we will focus exclusively on the epistemic implications of planetary contamination in the context of life detection, as these are the concerns that guide planetary protection policies [1]. This is not because moral risks are not deserving of consideration but rather that they are not currently constructively guiding action, while epistemic risks are the identified reasons for planetary protection as outlined above.

The epistemic risks of forward contamination originate from two distinct possibilities: (R1) bringing life to a dead planet fundamentally alters the dead planet, and (R2) bringing life to a living planet fundamentally alters the living planet.

For contamination of a dead planet (R1), the risk is only to the ascertainment of knowledge surrounding the natural trajectory of a planetary body's abiotic processes. We have already exposed the Martian environment to life as it is frequently assumed that some spores continue to survive on various landed Martian spacecraft, including the original Viking Landers [18–20]. Therefore the risk here is not whether a planet will be exposed to Earth life, but instead whether and how Earth life will persist on that planet [18,21]. If Earth life can rapidly embed itself in exotic abiotic environments, it could make it harder to understand the planet's native, pre-contaminated state. There may even be concern about contaminating hypothesized "prebiotic worlds," a term frequently used to describe Titan[5] [22–24]. In those situations one could imagine that the planet was naturally on a trajectory towards an independent emergence of life, but that our contamination causes “the ladder to be pulled up” [25,26] on nascent life. For all that we don’t know about life and how to detect it, we know even less about almost-life and how to detect it.[6]

[5] Likely because scientists often mistakenly treat organic and prebiotic chemistry as synonymous.

Contamination could also obscure our discovery or understanding of any extinct biospheres, via Earth life overwriting any remaining evidence of such biospheres. Although it is hard to imagine such overwriting happening catastrophically, given that Earth’s continuous global-scale geodynamicism and richly active biosphere have not prevented us from discovering evidence of life, or building a paleogeological understanding of Earth, throughout our biosphere’s history.

For contamination of a living planet, there is no additional risk of “pulling the ladder up” by the contaminating life, because presumably the ladder would have already been pulled up by native life[7]. The canonical concern is simply that Earth life would obliterate the native life, or alter and suppress it in such a way that Earth life ends up dominating the extraterrestrial biosphere. We can also imagine Earth life and native life evolve or merge into a new LUCA.

Finally there’s the possibility that Earth life and native life are not actually significantly different, either because life’s chemical organization is more deterministic than we might imagine (e.g., DNA is the only viable information encoding system; the canonical amino acids were selected for specific reasons; and protein folds used by Earth life are universal), or because other native life we discover in the solar system did not originate independently of Earth life (i.e., at some point(s) throughout the history of the solar system biological material was transferred between bodies via “natural”, non-intelligent means[8]). Should we contaminate a planet, resolving between these hypotheses would appear to become more difficult. Thus contamination in those scenarios risks our access to knowledge about the deterministic nature of life's biochemical details. However, we may be able to differentiate these possibilities using phylogenetic approaches as organisms originating from Earth (e.g., the contaminates) should root later in a phylogenetic tree than Martian organisms, which should appear to diverge from Earth life deep in life’s history.

## Our Ignorance is Nearly Complete

As with most topics in astrobiology, it is difficult to reason clearly about these epistemic risks, and how they should be translated into planetary protection recommendations, because of our profound ignorance about life’s origin and its nature beyond Earth. The most obvious aspect of this ignorance is that we have no way of assessing the likelihood of life emerging, either in general, or in specific planetary environments [27–31]. Presently, this widely discussed problem has no obvious solution. Most positions in this debate simply boil down to aesthetic preferences, unjustified heuristics or analogies[9]. More recent analysis indicates that life's early emergence on Earth *may* constitute evidence of its rapid emergence in analogous environments, but this is subject to interpretation of paleobiological evidence to date life's emergence, and the signatures of early life are strikingly similar to abiotic chemistry [32,33].

[6] What would be different between observations of “normal” abiotic chemistry that seems poised to remain indefinitely abiotic, and abiotic chemistry which is fated to become biological?

[7] A possible exception to this could be a living planet which contains multiple mostly-mutually exclusive biospheres à la shadow biospheres. A bit more on that later.

[8] In the literature this process is often called “lithopanspermia”, and if it is realistically viable, it’s even possible that life on Earth originated on a different planetary body [65].

However little we know about the origin of life, we know even less about how different biospheres could, or would interact. Planetary protection policies assume that sufficient delivery of Earth's biological material could completely consume another world's biosphere, and so it would follow that Earth life could possibly be completely overrun by an alien microbial system. Meanwhile there are legitimate questions about whether or not there are additional biospheres here on Earth—so called "shadow biospheres"—which coexist with us (the descents of LUCA), possibly through physical isolation, orthogonal biochemical processes, or dynamic (but difficult to detect) stable equilibria [34]. While neither biosphere subsumption nor shadow biosphere coexistence have been systematically tested, they seem to stand in complete contradiction to one another. Do we live in a reality where distinct biospheres can coexist? Or one in which their interaction must lead to destruction of one by way of the other?

As scientists and policy makers confront our profound ignorance related to the processes underlying planetary protection concerns, it is understandable to adopt conservative approaches, and assume the worst case scenario where evidence does not suggest alternatives. However, this approach, while understandable, is not a strictly *rational* way to balance our interest in exploration with our concerns of contamination [35]. A framework for making rational decisions requires first specifying the probability of different outcomes based on decisions, and the relative benefit or cost of those decisions [35]. As with most important or relevant decisions, we currently do not have good models of either the likelihood of different outcomes, nor their value. The question of the relative value (or cost) of different outcomes as it pertains to planetary protection is a question that must be addressed by society at large. But the question of the relative likelihood of various outcomes is something that we can reason about as scientists, even in the face of our profound uncertainty [36,37].

We examine the logic underpinning the fear reflexively evoked by planetary contamination, and its potential effect on our ability to detect native extraterrestrial life. Despite the layers of our uncertainty, we find that making statistically logical assumptions has the potential to free us from planetary protection guidelines which are loosely justified using historic, rather than scientific reasons, and replace them with a more permissive approach to planetary exploration which is grounded in our contemporary understanding of the nature of life.

---

[9] Consider for example the heuristic espoused by former president Obama about aliens, "they're real but I haven't seen them."[60] He later clarified that, "Statistically, the universe is so vast that the odds are good there's life out there." [61] Or the analogy from Fred Hoyle that, "The chance that higher life forms might have emerged in this way is comparable with the chance that a tornado sweeping through a junk-yard might assemble a Boeing 747 from the materials therein" [62]. Or a pithy summation from Scottish philosopher Thomas Carlyle who exclaimed, "A sad spectacle! If they be inhabited, what a scope for misery and folly; if they be na inhabited, what a waste of space!'' [63] For a detailed discussion of these different aesthetic arguments see the recording of the "Life Beyond Earth and Mind of Man" meeting from 1975 [64]. Or see the discussion from Davies of how the expectations of life have changed in the last century [34].

***Our near-complete ignorance on the likelihood of abiogenesis***

To start qualifying the risks inherent in the epistemic fears described above, we must first acknowledge that our ignorance on the likelihood of abiogenesis is nearly complete, and understand what that means for how we should determine our prior expectation of its emergence. Kite [27] and Kipping & Lewis [38] make the argument that we are completely ignorant even insofar as the order of magnitude of the likelihood of the origin of life [27,38]. This can be seen formally if we model the emergence of life on a planet as a binary variable (either life emerges or it does not), drawn from a Bernoulli distribution with a parameter $p$, which quantifies the likelihood of life's emergence [28,38]. If we know nothing about this process the uninformative prior follows a Beta distribution with parameters $\alpha = \beta = 0.5$ [39]. This distribution (**Fig. 1, left**) has a u-shape with the overwhelming majority of probability mass concentrated at the extreme ends of the distribution ($p \approx 0$ and $p \approx 1$).

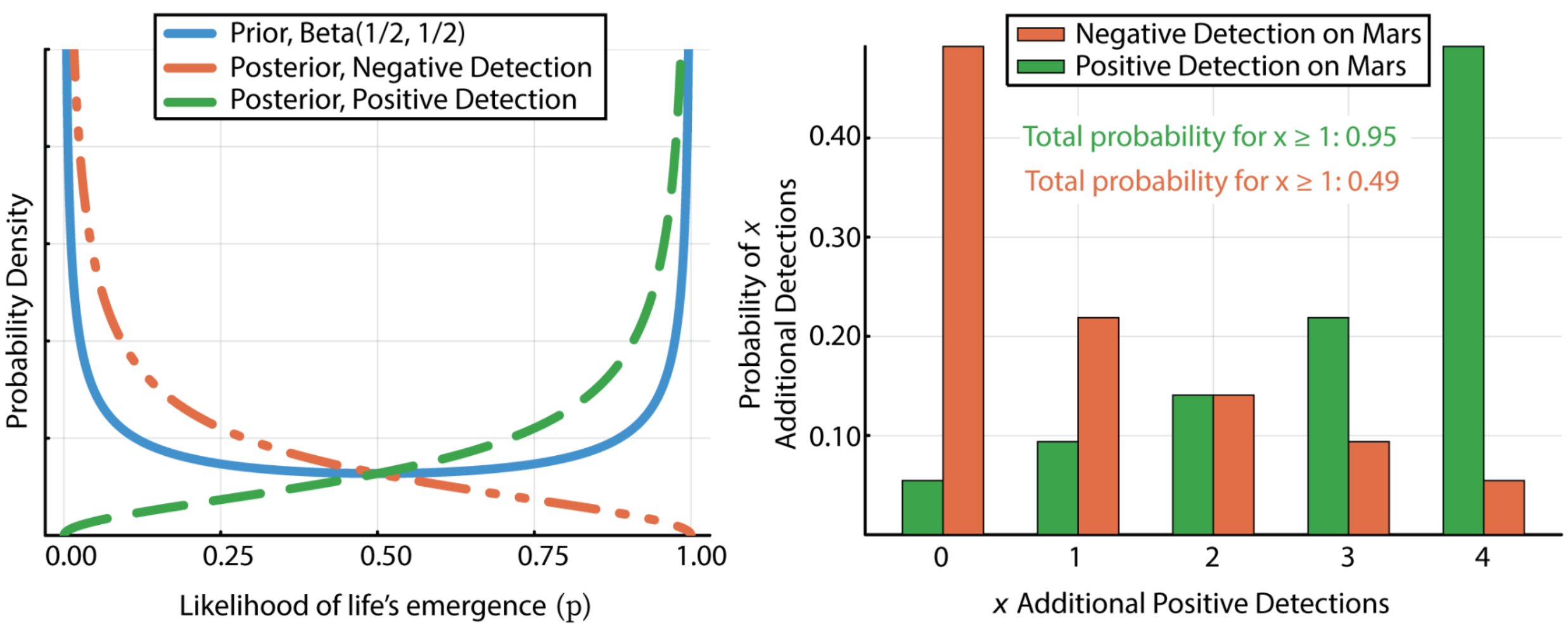

**Figure 1. Left.** The probability distribution over possible values of $p$, the likelihood that life emerges on a planet without any additional information. Our uninformed prior (blue line, modeled as a Beta distribution with shape parameters (0.5, 0.5)) indicates that the probability density space of abiogenesis is concentrated in the extremes—life should either be exceptionally unlikely, or exceptionally likely. The orange/green lines show the posterior after observing a single planet (beyond Earth) and concluding it is without/with life (negative/positive detection, respectfully). The posterior obtained after detection of only a single planetary target significantly shifts the probability density of abiogenesis such that it becomes either ubiquitous or vanishingly unlikely. **Right.** This can be seen by looking at the probability of detecting life on $x$ of 4 other targets observed. If we made a positive detection of life on Mars, then the most likely probability is that 4 of 4 additional targets observed will also have positive detections, and the least likely probability is that 0 of 4 additional targets will have positive detections. The result is reversed in the event that there is a negative detection[10] of life on Mars. We use Mars as an exemplar, but this logic would apply equally should our first detection be of another solar system body.

[10] By *negative* detection we mean ruling out the possibility that the planet was ever inhabited—in other words it implies proving the planet is genuinely abiotic. This is perhaps not possible, but it leaves the door open for research meant to understand how our confidence in the absence or presence of life impacts these probabilities.

This formulation captures the intuition clearly exemplified in a thought experiment about beakers from Haldane, and later by Jaynes [38,40,41]. Suppose you have a series of beakers filled with a volume of water set up in a fume hood with approximately the same conditions for each beaker. If you add compound X to each beaker, in what fraction of beakers should we expect the compound to be completely dissolved? Jaynes argues that if you know nothing else about the compound X (i.e., are ignorant about its fundamental properties), you should assume either approximately all the beakers will completely dissolve the compound or none of them will [40]. The reason is that there are simply many more ways for the compound to either completely dissolve or be completely insoluble, than there are ways for the compound to be just soluble enough to completely dissolve in some beakers but not in others. This is a consequence of the fact that completely dissolving the compound will depend on a large number of properties of the compound X, and their relation to the conditions in the water (the temperature, pressure, etc.). The chance that the properties of water somehow converge with the precise ways in which the conditions in our beakers differ (in the absence of any other information) would be exceptionally rare. Translating this to our understanding of the origin and distribution of life, it means that, absent any conclusive information on what makes life emerge on a planet, we should assume life in the universe is either extremely common, or extremely rare. Using similar reasoning to the beaker thought experiment, it would be unlikely that the universe is fine-tuned for life to emerge on planets that share environmental conditions (such as water activity, or water-rock interfaces) with Earth, but not to be ubiquitous given those conditions[11].

Based on these arguments, it would be very unlikely that life is just common enough to occur on Mars but no other "habitable"[12] environments in the solar system. Staying with the Bernoulli model outlined above—if we find evidence of life on Mars (past or present), we would update our uninformed prior based on detecting life on the first planet we searched. This updating is called Bayesian inference, and it works by combining a prior belief with observed data to produce a posterior—an updated distribution that reflects the newly observed evidence. Here, detecting life on Mars would shift the distribution of $p$ strongly toward higher values, ruling out the large region of parameter space where life is very rare. A negative detection would do the opposite. From this updated distribution we can predict the expected number of detections across future searches, with the spread of predictions reflecting our remaining uncertainty about $p$.[13]

If we assume that we will search for life on four additional targets in the solar system[14], the posterior resulting from a positive detection would give us confidence in detecting life on at least one additional target with a range of 67%-99.99999999%, and a mean confidence of 95% (**Fig. 1, right**). This happens mathematically because the detection of a new biosphere has effectively ruled out the parameter space where p is relatively small, which was a large fraction of the parameter space in the uninformed prior. This calculation excludes Earth life as evidence of life's likely emergence (as this can be argued to be anthropically biased). If we include Terran[15] life as evidence the confidence range changes to 90%-99.999999999[...]% chance of detecting further life in the solar system.

[11] Knowing which commonalities are important for abiogenesis is at present largely subjective. If we come to a better understanding of the role different features of planets play in this process, or if there's a difference in opinion to what we write here, the resulting posteriors would change.
[12] Where habitable can mean whatever one wants it to mean, as long as that belief is applied consistently to a candidate set of planets for life detection.
[13] Many good resources exist for learning about Bayesian inference, see e.g. [72,73].
[14] For example Venus, Europa, Enceladus, and Titan are often posited as life detection targets.

***Our near-complete ignorance on the susceptibility of planets to contamination from extraplanetary life***

The other major source of uncertainty we face is the ease at which biospheres interact with other biospheres, or with non-native geospheres. This is not a question of how often material is exchanged between solar system bodies, or even about the probability of life successfully hitching a ride on material in such a way that it can survive being ejected, transported across space, and land while maintaining viability[16]. The question is: presuming material exchange can safely deliver viable organisms from one planet to another, can such organisms then thrive and evolve on their adopted homes? We're concerned with the susceptibility of one planet to *invasion* from life of another planet, *given* that alien organisms make it safely to that planet. On Earth, problems of this sort are the central questions of *invasion biology* [42], and yet very little of the theory around planetary protection is informed by that discipline [12,43]. While most of our fears around forward contamination are based on horror stories of invasive species, evidence suggests that ecological invasions are the exception, not the rule [13].

Our biosphere is an emergent property of our geosphere and we don't have strong evidence that the biosphere can persist as a decoupled entity [44]. That is, we have no evidence that Earth life could take root and persist on Mars (just as we have no evidence that it could not take root and persist on Mars). We can draw an analogy again to the same thought experiment outlined above, but instead of dissolving a compound in a beaker, we can consider whether a microbial population, when added to a new environment, will grow, or decay. If we know nothing else about the microbial population or the environment, we should assume that it will either grow rapidly, or decay rapidly, because the state of the world in which the population of the contaminating taxa is introduced at its exact carrying capacity is indeed vanishingly small.

In the case of Terran life's arrival into an abiotic world the concept of primary succession may be more relevant. In that case the question would be whether or not the biological material delivered to the surface is capable of colonizing a pristine environment. The functional properties of primary colonizers are not well understood, but at least in the case of desiccated, nutrient poor environments, the capacity to fix atmospheric gases into organic feedstocks appears to be a key requirement [45]. Primary successors are also believed to often be dispersed from the atmosphere, and to be well suited to survive within the atmosphere [45]. It is not clear if these organisms would be likely to survive standard sterilization procedures, persist on spacecraft surfaces, or fix atmospheric sources of nutrients considering the substantial differences between Earth and Mars's atmospheres.

[15] Terran refers to the known global biosphere on Earth and not a shadow biosphere we have not yet detected.
[16] Although this is an area of active research. See for example [67–70] and references therein.

But the presumed lack of a shared evolutionary history severely complicates the analogy to invasion biology should Martian life exist. Invasive species emerge by exploiting untapped ecological niches in existing ecosystems, often by leveraging resources provided by biological components of the host ecosystem [46]. However when biospheres interact the resources provided by the host ecosystem may be irrelevant or actively harmful to the newly arrived biology—if Mars life and Earth life do not share biochemistry, the odds of successful invasion from Earth life, excepting primary accession, should decrease, not increase. Community-level planetary-protection assessments have primarily framed forward contamination of Mars in terms of whether terrestrial microorganisms can survive and proliferate under Martian environmental conditions, presupposing that any potentially extant Martian life would have little ecological resistance to invasion by space-faring Earth microbes [8,47,48]. If that was really the case, then either previous missions to Mars likely already contaminated the surface, or the Martian surface was likely already contaminated by Earth life due to natural material exchange between the planets earlier in the history of the solar system. However if the Martian surface is not already colonized by Terran microbes delivered by natural or technological sources, why should we assume that hypothetical Martians' ecological resistance is just good enough to have protected them thus far, but fragile enough for them to be overcome by additional landers? In fact contemporary studies of invasion potential repeatedly identify that the biggest determinant of invasion capacity is the frequency and timing of delivery, and not the size of delivered population [49–51]. Perhaps the risk is the 31st mission, not the 31st spore.

***Our near-complete ignorance on the fragility of nascent biospheres***

We can use similar arguments to reason that it doesn't make sense to worry about contaminating nascent biospheres, or "prebiotic" worlds. Based on our uncertainty it would be very unlikely to encounter a biosphere that is on the cusp of existence, yet unstable to the point that it could be destroyed by local perturbation. The unknown here is the likelihood that planets rapidly progress from so-called "prebiotic chemistry" to stable biospheres. First, it is not clear that there is any real distinction between an abiotic world and a "prebiotic" world, because we do not know if it can be knowable that a planet is on the way from non-life to life prior to that life emerging. Setting aside this conceptual issue, if we assume there is some kind of prebiotic phase where planetary chemistry is right on the cusp of transitioning to biochemistry, observing a planet undergoing an abiotic-to-biotic transitionary stage should be much more rare than simply observing a planet with life. This is based on the same kind of fine-tuning arguments as Jayne's beakers, and on the fact that the prebiotic period of Earth's history—if it ever truly existed—has a duration with an upper bound of about 1Gy, corresponding to the difference between the age of the Earth, and the age of the oldest preserved materials on Earth (which seems like a striking coincidence if this was actually how long it took to emerge)[17] [31]. "Prebiotic planets" shouldn't be common unless biological planets are very common.

***If biospheres are common we get second chances; if they're not, then there's nothing to worry about protecting***

As a consequence of being able to coarse grain both the probability space for *abiogenesis* into exceptionally rare or exceptionally common, and the probability space for *planetary invasion* into exceptionally hard or exceptionally easy, we can partition the possibility space into four scenarios (**Fig. 2**). Stepping through these, we can observe that many of the epistemic risks which have plagued both planetary protection and life detection discussions simply can't exist. It boils down to this: If biospheres are common we get second chances, and if they're not, then we don't have to worry about protecting alien life in the solar system because it's not there. Below we walk through each entry in the possibility matrix, and discuss the implications in the context of Mars:

I. **Abiogenesis common; invasion easy**: Mars likely has a biosphere. It may or may not be independent of ours, but would have been both exposed to our biosphere on numerous occasions, as well as exposed our biosphere to its own on numerous occasions. It's hard to imagine spacecraft contamination obliterating life if there have been lots of exposures through natural material exchange and previous missions to the Martian surface .
II. **Abiogenesis common; invasion hard:** Mars likely has a biosphere. It is likely independent of ours. Contamination is unlikely to threaten native biospheres, and there will be many other chances to look for life in the solar system, and universe.
III. **Abiogenesis rare; invasion easy:** Mars may or may not have a biosphere, but if it does, it likely shares an origin with our own. The risk of catastrophe from contamination seems unlikely due to shared origins (even with invasion being easy), and contamination itself would carry less epistemic risk as we aren't compromising knowledge about a second origin of life.
IV. **Abiogenesis rare; invasion hard:** Mars likely does not have a biosphere, and so there is no epistemic risk to destroying what does not exist.

[17] This upper bound is unlikely to get unequivocally pushed back before we have an agreed upon theory of planets and life, simply because the evidence starts to become too contentious.

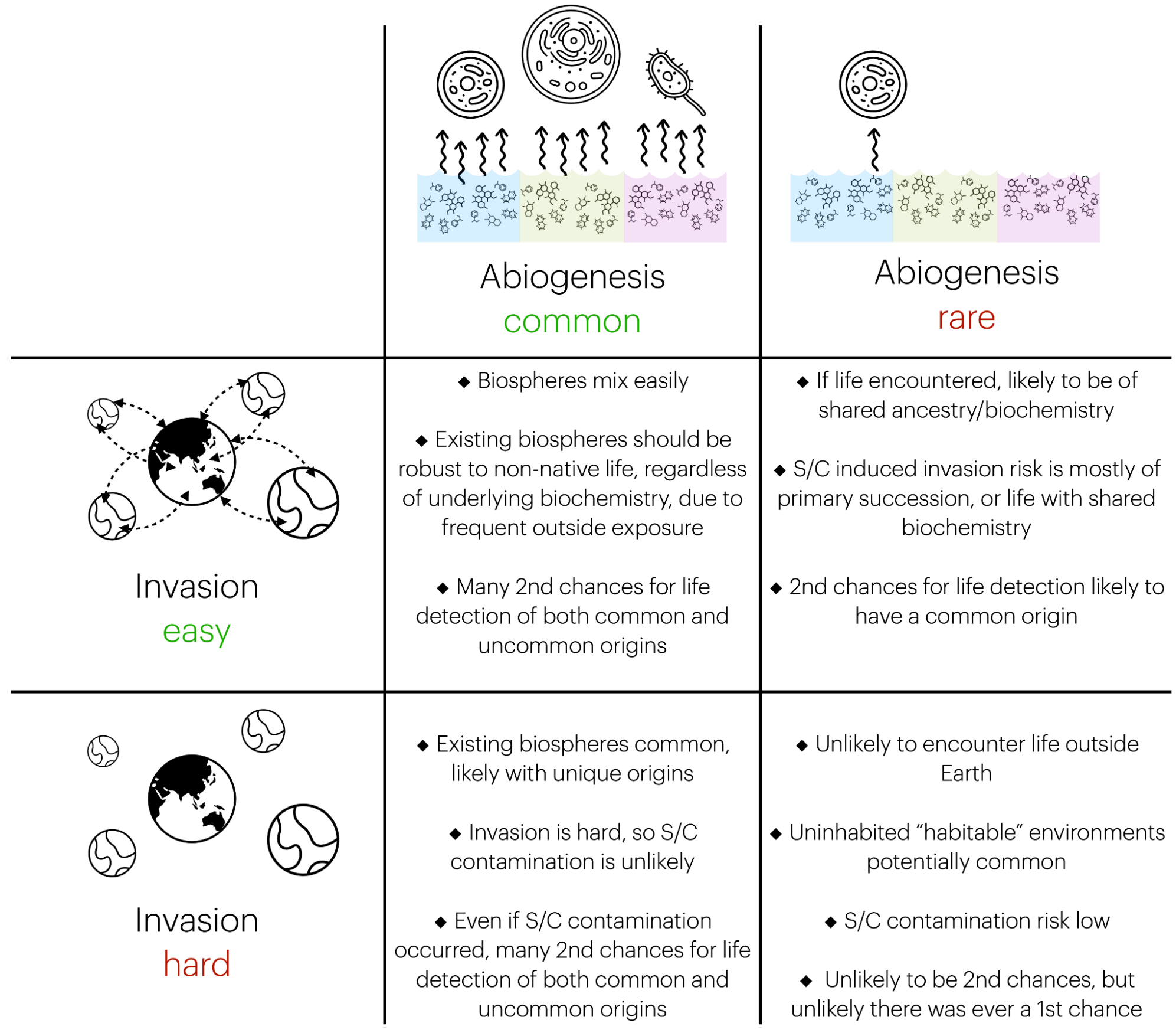

**Figure 2.** The matrix of possibilities resulting from being able to coarse grain both the probability space for *abiogenesis* into exceptionally rare or exceptionally common, and the probability space for *planetary invasion* into exceptionally hard or exceptionally easy. Here, we talk about a biosphere being "robust" to invasion or contamination similarly to how the term is used in evolutionary biology—that it can persist (does not collapse, and is not significantly altered) in the face of perturbation (contact from exogenous life).

So, why does Microbe 31 pose such a grave risk to Mars? It's not impossible that the 31st microbe on a mission to a special region of Mars could make it much more difficult to ascertain knowledge about a Martian biosphere. But only if Mars has (or had) a biosphere to begin with, *and* microbe 31 can grow and thrive in the Martian environment, *and* the Martian biosphere has no resistance to biological invasion, *and* the hundreds of thousands of previously delivered microbes from prior missions somehow did not lead to a catastrophic outcome *and* a complete obliteration of the Martian biosphere (or its fossils) is possible. Or perhaps microbe 31 poses no more of a risk than the 30th microbe, or the 300,000th microbe.

## The consequences of our ignorance

***Clean rooms shouldn't be clean—We should develop adaptive sterilization methods***

As others have pointed out, any amount of space exploration will inevitably include the transportation of microorganisms from Earth to new planetary environments [11,19]. Current paradigms focus on sterilization for preventing forward contamination [11,21,52]. However, even with the most stringent protocols for clean rooms, Terran life rapidly colonizes space craft and pristine samples, although it seems those populations may not ultimately stabilize to detectable abundances in the long run [11,52–54]. As we have discussed, the locus of our uncertainty is not whether or not organisms will survive the trip to other planets (evidence seems to point to this being within the realm of the probable). Instead, it is about whether those organisms will thrive, adapt, and compete with (possible) indigenous life in other environments. Given this, we propose that planetary protection procedures focus on mitigating the evolutionary and metabolic potential of spacecraft, rather than on pure sterilization. For example, we believe the community can draw insight from adaptive therapies for cancer treatment [55].

The goal should be to create an environment on spacecraft that selects for organisms which will be poorly adapted to either the journey to a new planet (including the extreme radiation, thermal, and pressure environments), poorly suited to the novel planetary conditions (including extreme resource limitation), or otherwise incapable of sustaining a sufficiently diverse population to adapt to novel conditions (potentially by limiting their evolutionary potential through metabolic constraints, small effective population sizes, or with limited capacity to form phenotypically diverse, genotypically identical populations). Current sterilization procedures are likely driving adaptation to nutrient limitation and resilience to extreme conditions, which may paradoxically facilitate the contamination of alien worlds [56]. A consequence of this approach would be that flight-ready "clean rooms," may need to be densely populated with microorganisms, or well-suited to non-extremophiles, in order to ensure that the majority of the bioburden sent to other worlds is composed of lineages poorly suited to adapt to novel environments.

***We should search for shadow biospheres if we might eradicate the obvious one***

One of the proposed risks includes the possibility that Earth life coexists or competes with indigenous life on other worlds. In that case the epistemic concern is that we will be unable to detect or learn about the indigenous life as it may be overwritten by Terran life processes. However, unless that overwriting is complete there will still be signatures of indigenous life on a planet. In that case the challenge is not how to prevent Terran life from arriving in a new environment, but instead on how to detect a biosphere in the context of existing Terran life. This is the same problem as searching for a shadow biosphere here on Earth [34]. Many concerns around planetary protection would be alleviated if methods for detecting shadow biospheres were developed. This is something that can be pursued concurrent with further space exploration, and could have a profound positive impact on the state of our knowledge both about how biospheres emerge and how they interact. If we can definitively determine there are no other biospheres on Earth, we will have further evidence that life either "pulls the ladder up" on other biospheres through competitive exclusion, or that the emergence or persistence of shadow life is very rare given contemporary Earth conditions.

## Discussion

We have argued that current concerns over forward contamination of planetary bodies in the solar system may be overblown, and even if they are not, our approach to managing them may be misplaced. We have argued that further research in detecting shadow biospheres on Earth may reduce the downside consequences of forward contamination, and we have discussed how adaptive sterilization procedures may be more effective at managing the risk of forward contamination in the first place.

The core of our argument is that we have essentially no reason to believe that either the origin of life in the solar system is exceptionally easy, or exceptionally hard. We illustrated the consequences of this by modeling the likelihood of life's emergence as a Bernoulli variable with an unknown probability $p$. This showed that rational priors over $p$ would lead us to conclude that either: Mars is inhabited, and we likely have other biospheres in the solar system; or Mars is uninhabited, and we likely have no other biospheres in the solar system. We made an argument following the same statistical logic that either: we would expect the ecological resistance of alien biospheres to nearly universally prevent invasion from Terran life, in which case the risks of forward contamination are small; or alien biospheres may be exceptionally susceptible to invasion, in which case our current activity on Mars has likely already altered any potential Martian biosphere—implying that the horse has already left the barn and our stringent approach to planetary protection is only limiting our knowledge of Martian life, not protecting it. As we have described, our analysis is framed in the context of Mars exploration, but our argument is largely independent of which solar system environments are explored first. We used Mars as a narratively convenient example, since the majority of the planetary protection and life detection discussions throughout history have centered on Mars.

However, this "beaker model" entails assumptions about the likelihood of life's emergence and the interactions of microbial species. In the case of life's emergence it assumes all planets deemed "habitable" are "equal" beakers, but this might not be the case and might be dependent on relative frequency of planetary properties, such as: the availability of certain elements or molecules, the atmospheric or subsurface temperature or pressure, the volume of solvents such as water or hydrocarbons, or the surface area of such solvents in contact with solid rock or ice. If planetary heterogeneity significantly and complicatedly modulates the likelihood or possibility for life to emerge, then our calculations would be inapplicable as the likelihood of life emerging in each planetary environment would not be easily reducible to simple bulk characteristics.. In the case of ecological resistance to alien microbes, this formalization assumes that all microbes interact with all alien environments in a similar manner. It ignores specific ecological considerations that contribute to invasive potential, and coarse grains the complexity of ecological invasion into a binary probability. It is certainly the case that on Earth these processes exhibit exceptional interspecific, intraspecific, and temporal variability, not to mention significant abiotic factors [13,49,50]. Nevertheless, these nuanced distinctions are likely eclipsed by the truly alien environment presented by other worlds. Importantly, the arguments we have laid out here are about the risks of robotic exploration, where some level of sterilization (adaptive or traditional) is possible, and the mission itself does not bring the diversity and density of life that would be inevitable for human exploration.

Concerns about planetary protection, as with definitions of life, are implicitly shaped by theories of life [57]. But these theories of life are rarely stated explicitly even when being used to justify planetary protection protocols. Our argument is not that planetary protection considerations do not matter. Rather, it is that there is no clear strategy to come to understand the factors which would ultimately determine the risks of planetary exploration. One possibility is that a rigorous evaluation of the likelihood of various outcomes—whether that be the likelihood of life on other planets, or the likelihood of planetary contamination—may be impossible before we identify or create a novel biosphere [29]. Ultimately, doing science is premised on interacting with the world in order to gain understanding of how it works. This is not a risk free enterprise, but it's the only approach if we admit that the nature of reality is not deducible from pure thought alone. If planetary protection protocols are so severe as to hamper our ability to conduct experiments, they are in a sense self-undermining. Current policies poorly reflect what we know about the trade-offs in the face of this tension between exploration and protection.

Right now, "special regions" on Mars, and other environments in the solar system with liquid water (either presumed or certain) are subject to strict planetary protection guidelines. The consequence of this is that we are trying to detect life as definitively as possible, while simultaneously trying to avoid environments in the solar system which give us the greatest chances of detecting life. There is clearly an oxymoronic tension in these efforts. We propose, based on the arguments in this paper, that the risks of contamination of these "special regions" is vastly outweighed by the potentially decisive information they could provide about life in the universe. We believe that resisting closure to the question [58] of whether or not life exists on Mars, be it positive or negative, isn't just counterproductive, but unjustifiable based on planetary protection concerns. We don't believe the risks from Microbe 31 should prevent us from using the most decisive and obvious strategies to search for extraterrestrial life.

As with most issues in astrobiology, disagreements about planetary protection policies cleave around debates on the nature of life and the endless attempts to answer the question "what is life?" Is our biosphere a freak accident? A cairn of meticulously balanced stones, ready to topple when an intrusive pebble is carelessly introduced? Then we should take no risks of returning material to Earth as backward contamination is cataclysmic. Or is our biosphere an unstoppable causal force in the universe? A relentlessly expanding pattern in matter that can disrupt and assimilate any other pattern that crosses its path? In that case we can take no risks with forward contamination, as we are patient zero of an irreversible transition to Earth-likeness in the universe. Whatever we suppose about our biosphere, why should we assume it is different from others? What is the nature of interacting biospheres? Is the metaecology of their interaction "red in tooth and claw"? Or is the better analogy a collision of galaxies, where almost no matter actually collides? We may not know the true nature of life before we detect other biospheres, and we may not detect other biospheres if we recoil, without careful justification, when the most promising places to close the question present themselves.